\documentclass[12pt]{article}
\input psfig.sty

\def\xbe7{$^7Be$}
\def\be7pg{$^7Be(p,\gamma)^8B$}
\def\b8{$^8B$}
\def\n16{$^{16}N$}

\begin{document}

\title{
  Precision Measurements of the \be7pg Reaction \\
With Radioactive Beams \\
and the \b8 Solar Neutrino Flux 
\footnote{Work Supported by USDOE Grant No. DE-FG02-94ER40870.}}

\author
{Moshe Gai \\
 Laboratory for Nuclear Science, Dept. of Physics, U3046, \\ 
University of Connecticut, 
    2152 Hillside Rd., Storrs, CT 06269-3046, USA \\ 
   (gai@uconn.edu, http://www.phys.uconn.edu)}

\maketitle

\begin{abstract}
The \be7pg reaction is one of the major source of uncertainties 
in estimating the \b8 solar neutrino flux and is critical for  
undertsanding the Solar Neutrino Problem and neutrinos. 
The main source of uncertainty is the existence
of conflicting data with different absolute normalization.
Attempts to measure this reaction rate with \xbe7 beams are 
under way by the UConn-LLN collaboration, and we discuss a 
newly emerging method to extract this cross 
section from the Coulomb dissociation of the radioactive beam of \b8.
We discuss some of the issues relevant for this study including the 
question of the E2 contribution to the Coulomb dissociation process 
which was measured to be small.
The Coulomb dissociation appears to provide a viable alternative method 
for measuring the \be7pg reaction rate, with a weighted average of 
the RIKEN1, RIKEN2 and GSI1 published results of 
$ S_{17}(0) = 19.4 \pm 1.3 $ eV-b.

\end{abstract}

\section{Introduction: The \be7pg Reaction at Low Energies}

The solar neutrino problem \cite{Book,Phs96} may allow for new
break through in our understanding of the standard model~\cite{Wo78,Mi85}.
The precise knowledge of the astrophysical $S_{17}$-factor of the
$^7$Be(p,$\gamma$)$^8$B reaction
at the Gamow peak, about 20 keV, is of crucial importance
for interpreting terrestrial measurements of the solar neutrino
flux~\cite{Ade98}.
This is particularly true for the interpretation of results from the
Homestake, Kamiokande, SuperKamiokande and SNO experiments
~\cite{Phs96} which
measured high energy solar neutrinos mainly or solely from $^8$B decay.

Eight direct measurements were reported for the $^7$Be(p,$\gamma$)$^8$B reaction
~\cite{Kavanagh60,Parker68,Kavanagh69,Vaughn70,Wiezorek77,Filippone83,Hammache98,
Ha99}.
Since the cross section at $E_{\rm cm} \approx$~20~keV
is too small to be measured,
the $S_{17}$-factors at low energies have to be extrapolated
from the experimental
data with the help of theoretical models (see e.g. Johnson
{\it et al.}~\cite{Johnson92},
Jennings {\it et al.}~\cite{Jennings98}).
But also in the energy range above $E_{\rm cm} \approx$~100~keV,
where measurements can be performed, experimental difficulties,
mainly connected with the determination of the effective target
thickness of the radioactive $^7$Be target,
hamper the derivation of accurate results.
These difficulties are reflected in the fact that the six precise
measurements~\cite{Parker68,Kavanagh69,Vaughn70,Filippone83,Hammache98, Ha99}
can be grouped into two distinct data sets which agree in their
energy dependence but disagree in their absolute normalization
by about 30\%.
Since this discrepancy is larger than the error of adopted $S_{17}$
in the standard solar model,
experimental studies with different methods are highly desirable
for improving the reliability of the input to the standard solar
model~\cite{Bahcall92}.

In Fig. 1 we show the world data including our new GSI measurement of
$S_{17}$ \cite{Iw99}.  It is clear from this
figure that if we are to quote $S_{17}(0)$
with an accuracy of $\pm 5\%$, many more experiments using different
methods are required.

\centerline{\psfig{figure=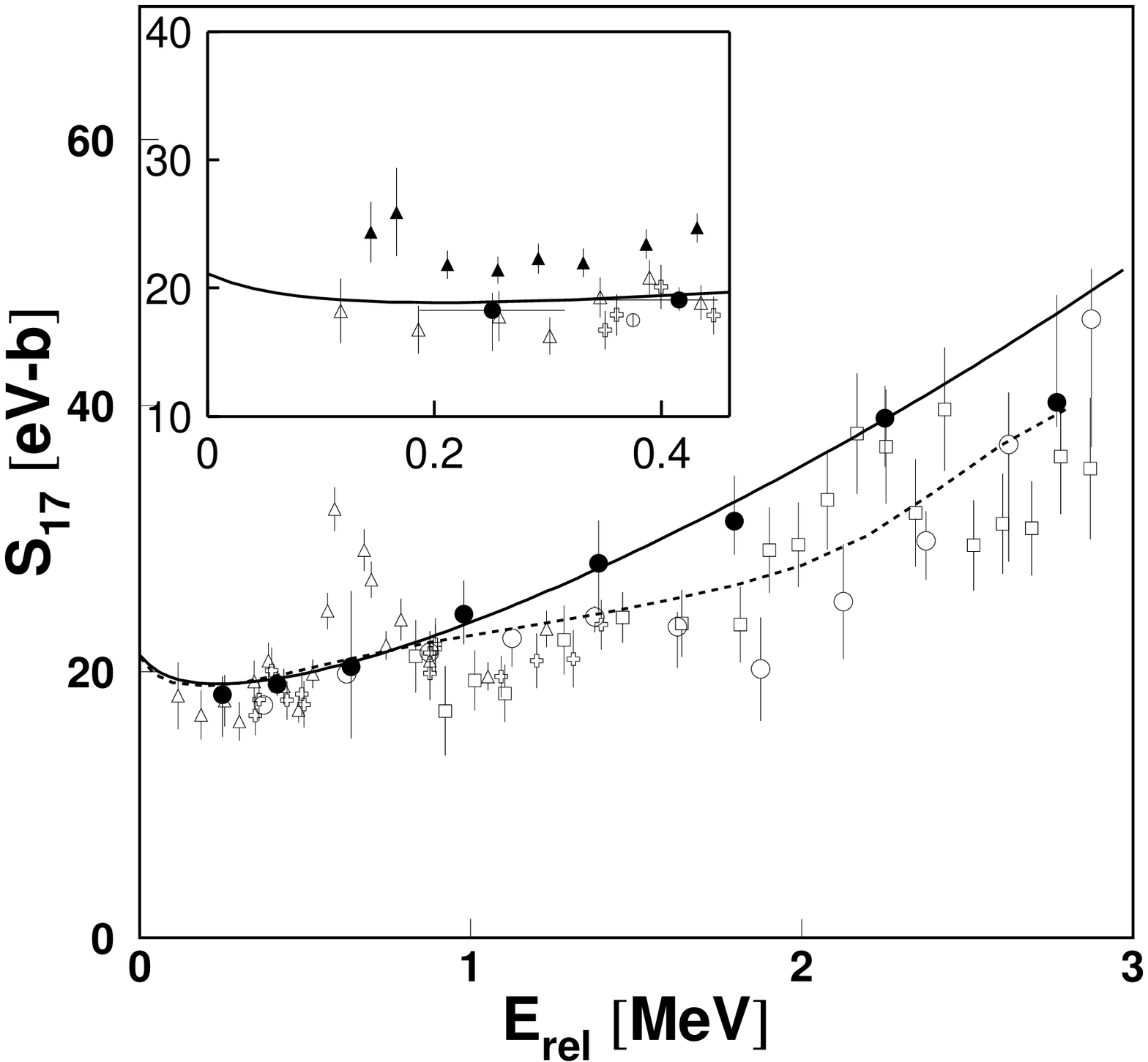,height=4in}}

\hspace{0.5in} \underline{Fig. 1:}  World data on $S_{17}$ as reported
       by our GSI collabortion \cite{Iw99}.

\subsection{Extrapolation Procedures}

The discrepancy in the measured absolute value of the
cross section of the \be7pg reaction is quite possibly best addressed with
a $^7Be$ radioactive beam and a hydrogen target, allowing for a
direct measurement of the beam-target luminosity as we propose
to do at LLN. However, additional
uncertainty exists in the theoretical extrapolation of the measured
cross section to solar energies (approximately 20 keV).
A few theoretical studies suggest an extrapolation procedure that
is accurate to approximately $\pm 1\%$
\cite{Jennings98}.  Without discussing these
rather strong statements we consider a similar situation that haunted
Nuclear Astrophysics a few years back-- the S-factor of the $d(d,\gamma)^4He$
reaction. It was assumed that in this case d-waves dominate and no
nuclear structure effects should play a role at very low energy,
as low as 100 keV.
Much in the same way, it is stated today that s-waves dominate the
\be7pg reaction and we do not expect nuclear structure effects to
play a role at low energies in the
\be7pg reaction. In Fig. 2 we show Fowler's extrapolated
d-wave S-factor that is a mere factor of 32 smaller than measured, due to a
small non d-wave component in the d + d interaction \cite{Ba87}. A small
nuclear structure effect, namely the d-wave component of the ground
state of $^4He$, gives rise to a change by a factor of 32 in the predicted
astrophysical S-factor. Similarly we may ask whether a small 
non s-wave component in the low energy interaction of p + \xbe7
could alter the extrapolated $S_{17}(0)$ value by considerably more than
one percent. A measurement of $S_{17}(0)$ with an accuracy of $\pm 5\%$
mandates that the cross section be measured at low energies, as low as
possible, so as to also test the extrapolation procedures.

\centerline{\psfig{figure=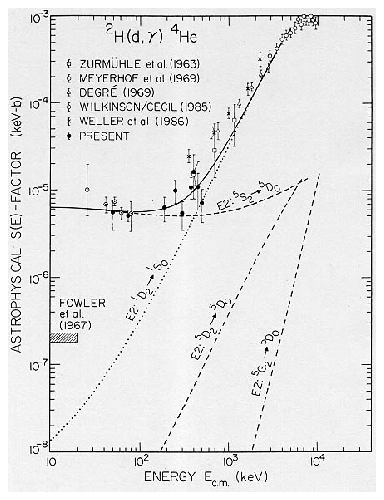,height=4in}}

\hspace{0.5in} \underline{Fig. 2:}  Extrapolation of d-wave S-factor
    of the $d(d,\gamma)^4He$ reaction\cite{Ba87}. Note the presence of
    small non d-wave components that yield a discrepancy from Fowler's
    extracted S-factor by a factor of 32.

\section{The Coulomb Dissociation of $^8B$ and the \be7pg Reaction
     at Low Energies}

The Coulomb Dissociation (CD) \cite{Bau86,Be88} is a
Primakoff \cite{Pr51} process that could be viewed in first order as the time
reverse of the radiative capture reaction.  In this case instead of studying
for example the fusion of a proton plus a nucleus (A-1), one studies the
disintegration of the final nucleus (A) in the Coulomb field, to a proton
plus the (A-1) nucleus.  The reaction is made possible by the absorption of
a virtual photon from the field of a high Z nucleus such as $^{208}Pb$.  In
this
case since $\pi/k^2$  for a photon is approximately 1000 times larger
than that
of a particle beam, the small cross section is enhanced.  The large virtual
photon flux (typically 100-1000 photons per collision) also gives rise to
enhancement of the cross section.  Our understanding of the Coulomb
dissociation process \cite{Bau86,Be88} allows us to extract the inverse
nuclear process even when it is very small.  However in Coulomb
dissociation since $\alpha Z$  approaches unity (unlike the case in electron
scattering), higher order Coulomb effects (Coulomb post acceleration) may
be non-negligible and they need to be understood
\cite{Ber94,Typ94}.  The success of CD experiments \cite{Ga95} is in
fact contingent on understanding such effects and
designing the kinematical conditions so as to minimize such effects.

Hence the Coulomb dissociation process has to be measured with great care
with kinematical conditions carefully adjusted so as to minimize nuclear
interactions (i.e. distance of closest approach considerably larger then 20
fm, or very small forward angles scattering), and measurements must be
carried out at high enough energies (many tens of MeV/u) so as to maximize
the virtual photon flux.

\subsection{The RIKENII Results at 50 AMeV}

The RIKENII measurement~\cite{Ki97} of detailed angular
distributions for the Coulomb dissociation of \b8 allowed us
to extract the E2 amplitude in the CD of \b8.
The same data also allowed us to extract values for $S_{17}$
as recently published \cite{Ki98}.
The $^{208}$Pb target and $^8$B beam properties
in this experiment were as in Ref. \cite{Mo94},
but the detector system covered a large angular range up to around 9$^{\circ}$
to be sensitive to the E2 amplitude.
The E1 and E2 virtual photon fluxes were calculated
\cite{Ki97} using quantum mechanical approach.
The nuclear amplitude is evaluated based on the collective form factor
where the deformation length is taken to be the same as the Coulomb one.
This nuclear contribution results in possible uncertainties in the fitted
E2 amplitude. Nevertheless, the present results lead to a very small
E2 component at low energies, below 1.5 MeV, of the order of a few
percent, even smaller than the low value predicted by Typel and
Baur \cite{Typ94} and considerably below the upper limit on the E2
component extracted by Gai and Bertulani
\cite{Ga94} from the RIKENI data.
These low values of the E2 amplitude were however recently challenged
by the MSU group \cite{Davids} that used interference between E1 and E2
components to observe an asymmetry of the longitudinal-momentum
distribution of $^7$Be \cite{Esben}. They \cite {Davids} claim:
$S_{\rm E2}$/$S_{\rm E1}=6.7^{+2.8}_{-1.9}\times 10^{-4}$ at 0.63 MeV.
These results are considerably larger than reported by Kikuchi {\em et al.}
\cite{Ki97}, and just below the upper limit reported by Gai and
Bertulani \cite{Ga94}. A recent analysis
of the RIKENII data \cite{Ki97}
by Bertulni and Gai \cite{Ber98} confirmed the
small E2 extracted by Kikuchi et al. \cite{Ki97} as well as
the negligible nuclear contribution. Note that in this analysis
\cite{Ber98} the acceptance of the RIKENII detector is taken into
account using the matrix generated by Kikuchi {\em et al.} \cite{Ki97}.
Recently a possible mechanism to reduce the E2 dissociation amplitude
was proposed by Esbensen and Bertsch \cite{Esben}.

\subsection{The GSI Results at 254 A MeV}

An experiment to measure the Coulomb dissociation of $^8$B at a
higher energy of 254~$A$~MeV was performed at GSI \cite{Iw99}.
The present experimental conditions have several advantages:
(i) forward focusing allows us to use the magnetic
spectrometer KaoS~\cite{Senger93} at GSI for
a kinematically complete measurement with high detection
efficiency over a wider range of the p-$^7$Be relative energy;
(ii) because of the smaller influence of straggling on the
experimental resolution at the higher energy, a thicker target
can be used for compensating the weaker beam intensity, (iii)
effects that obscure the contribution of E1 multipolarity to
the Coulomb dissociation like E2 admixtures and higher-order
contributions are reduced~\cite{Ber94,Typ94}. The contribution
of M1 multipolarity is expected to be enhanced at the higher energy,
but this allows to observe the M1 resonance peak and determine its
$\gamma$ width.

A $^8$B beam was produced by fragmentation of a 350~$A$~MeV $^{12}$C
beam from the SIS synchrotron at GSI that impinged on a beryllium
target with a thickness of 8.01 g/cm$^2$.
The beam was isotopically separated by the fragment separator
(FRS)~\cite{Geissel92}
by using an aluminum degrader with a thickness of 1.46 g/cm$^2$
with a wedge angle of 3 mrad.
The beam was transported to the standard target-position of
the spectrometer KaoS~\cite{Senger93}.
The average beam energy of $^8$B in front of the breakup
target was 254.5~$A$~MeV,
a typical $^8$B intensity was 10$^4$ /spill (7s/spill).
Beam-particle identification was achieved event by event with
the TOF-$\Delta E$ method by using a beam-line plastic scintillator
with a thickness of 5 mm placed
68 m upstream from the target and a large-area scintillator wall
discussed later placed close to the focal plane of KaoS.
About 20 \% of the beam particles were $^7$Be, which could
however unambiguously be
discriminated from breakup $^7$Be particles by their time of flight.

An enriched $^{208}$Pb target with a thickness of
199.7 ($\pm$ 0.2) mg/cm$^2$ was placed at the entrance of KaoS.
The average energy at the center of the target amounted
to 254.0~$A$~MeV. The reaction products, $^7$Be and proton,
were analyzed by the spectrometer
which has a large momentum acceptance of $\Delta p/p \approx 50$~\% and
an angular acceptance of 140 and 280~mrad in horizontal and vertical
directions, respectively.
For scattering-angle measurement or track reconstruction of
the two reaction products,
two pairs of silicon micro-strip detectors were installed
at about 14 and 31~cm downstream from the target, respectively,
measuring either x- or y-position of the products before
entering the KaoS magnets.
Each strip detector had a thickness of 300~$\mu$m, an active
area of 56 $\times$ 56 mm$^2$, and a strip pitch of 0.1 mm.

The measured complete kinematics of the breakup products
allowed us to reconstruct the p-$^7$Be relative energy and
the scattering angle $\theta_8$ of
the center-of-mass of proton and $^7$Be (excited $^8$B) with
respect to the incoming beam from the individual momenta and
angles of the breakup products.

To evaluate the response of the detector system, Monte-Carlo
simulations were performed using the code GEANT\cite{GEANT}.
The simulations took into account
the measured $^8$B beam spread in energy, angle, and position
at the target, as well as the influence of angular and energy
straggling and energy loss in the layers of matter.
Losses of the products due to limited detector sizes were also accounted for.
Further corrections in the simulation are due to the feeding of
the excited state at 429~keV in $^7$Be.
We used the result by Kikuchi {\em et al.}~\cite{Ki97} who
measured the $\gamma$-decay in coincidence with Coulomb
dissociation of $^8$B at 51.9~$A$~MeV.

The Monte Carlo simulations yielded relative-energy resolutions
from the energy and angular resolutions of the detection system
to be 0.11 and 0.22~MeV (1$\sigma$) at $E_{\rm rel}=0.6$ and
1.8 MeV, respectively.
The total efficiency calculated by the simulation was
found to be larger than 50\% at $E_{\rm rel}= 0-2.5$~MeV.
due to the large acceptance of KaoS.

\centerline{\psfig{figure=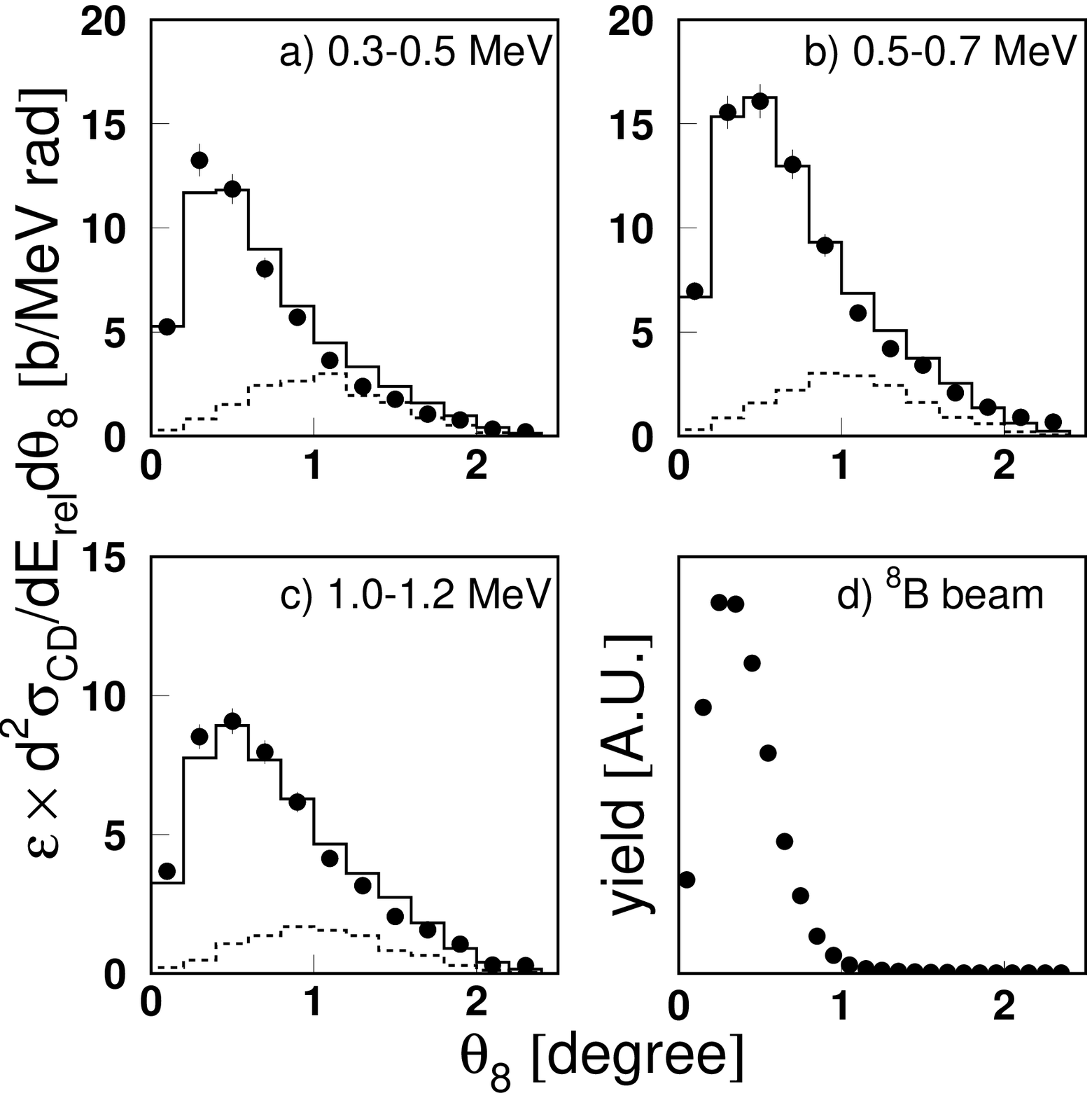,height=4in}}

\hspace{0.5in} \underline{Fig. 3:}  (a,b,c) Yields of breakup events plotted 
   against $\theta_8$, the scattering angle of the excited $^8$B, for three 
   relative energy bins.  The full histograms show the results of a simulation 
   taking into account the measured angular spread of the incident $^8$B beam 
   (shown in d) and assuming E1+M1 multipolarity.  The dashed histograms show 
   the simulated results for E2 contribution according to the calculations of 
   Bertulani and Gai\cite{Ber98}.

In Fig. 3, the experimental yield is plotted
against the scattering angle of the excited $^8$B for three
relative-energy bins,
0.3$-$0.5 (a), 0.5$-$0.7 (b), and 1.0$-$1.2 (c) MeV,
together with results of the Monte-Carlo simulations assuming E1 excitation.
The M1 transition also contributes to the 0.5$-$0.7 MeV bin
with the same angular dependence.
Since we did not measure the incident angle of $^8$B
at the target, the experimental angular-distributions
represent the $\theta_8$ distributions folded with the
angular spread of the incident beam shown in Fig. 3d.
The angular resolution was estimated to be 0.35$^\circ$ (1$\sigma$),
smaller than the observed widths.
As seen in Fig. 3
the experimental distributions are well reproduced by the
simulation using only E1 multipolarity, in line with the
results of Kikuchi {\it et al.}~\cite{Ki97}.

To study our sensitivity to a possible E2 contribution,
we have added the simulated E2
angular distribution from the E2 Coulomb dissociation
cross section calculated by Bertulani and Gai \cite{Ber98}.
Note that nuclear breakup effects are also included in
the calculation. By fitting the experimental angular
distributions to the simulated ones, we obtained 3$\sigma$
upper limits of the ratio of the E2- to E1-transition
amplitude of the $^7$Be(p,$\gamma$)$^8$B reaction,
$S_{\rm E2}$/$S_{\rm E1}$ of $0.06\times 10^{-4}$,
$0.3\times 10^{-4}$ and $0.6\times 10^{-4}$ for
$E_{\rm rel}=$ 0.3$-$0.5, 0.5$-$0.7 and 1.0$-$1.2 MeV,
respectively. These numbers agree well with the results of Kikuchi
{\it et al.} \cite{Ki97}, and suggest that the results of
the model dependent analysis of Davids {\em et al.} \cite{Davids} needs
to be checked.

\subsection{Conclusion: The Coulomb Dissociation Method}

In conclusion we demonstrated that the Coulomb dissociation (when
used with "sechel") provides
a viable alternative method for measuring small cross section of
interest for nuclear-astrophysics. First results on the CD of \b8
are consistent with
the lower measured values of the cross section and weighted 
average of all thus far data is: $S_{17}(0)\ = \ 19.4 \pm
1.3 \ eV-b$, see Table 1. The accuracy of the extracted S-factors are now
limited by our very understanding of the Coulomb dissociation process,
believed
to be approx. $\pm$10\%. The value of the E2 S-factor as extracted from both
the RIKEN and GSI experiments are consistent and shown to be very small,
$S_{E2}/S_{E1}$ of the order of $10^{-5}$ or smaller, see Table 1.

\begin{center}

\underline{Table 1:} Measured S-factors in Coulomb 
      dissociation experiments.
\end{center}

\begin{tabbing}

\hspace{1in} \= \underline{Experiment} 
\hspace{1in} \= \underline{$S_{17}(0)$ eV-b}
\hspace{1in} \= \underline{$S_{E2}/S_{E1}$(0.6 MeV)} \\
\   \\
\> RIKEN1 \> $16.9 \pm 3.2$ \> $< 7 \times 10^{-4}$ \\
\   \\
\> RIKEN2 \> $18.9 \pm 1.8$ \> $< 4 \times 10^{-5}$ \\
\    \\
\> GSI1 \> $20.6 \pm 1.2 \pm 1.0$ \> $< 3 \times 10^{-5}$ \\
\    \\
\> MSU \>                         \> $6.7\ + 2.8\ -1.9 \times 10^{-4} $\\
\    \\
\> \underline{ADOPTED} \> $19.4 \pm 1.3$ \> $< 3 \times 10^{-5}$ \\

\end{tabbing}

\section{The UConn-LLN Experiment With $^7Be$ Radioactive Beam}

We propose to measure the cross section of the \be7pg reaction with
the use of \xbe7 beams produced at Louvain-La-Neuve (LLN) \cite{Gai98}.
We intend to perform two experiments, both with \xbe7 beams
at Louvain-La-Neuve.  The first one is already approved 
(PH120) with the use of the old
cyclotron for three data points measured at 1.0-0.5 MeV. In a
second experiment we propose to
use the new cyclotron at LLN for additional measurements below 0.5 MeV.
It is however clear that a test of the extrapolation procedures,
as discussed above, will require yet a third experiment with an
implanted \xbe7 and low energy proton beams.

\centerline{\psfig{figure=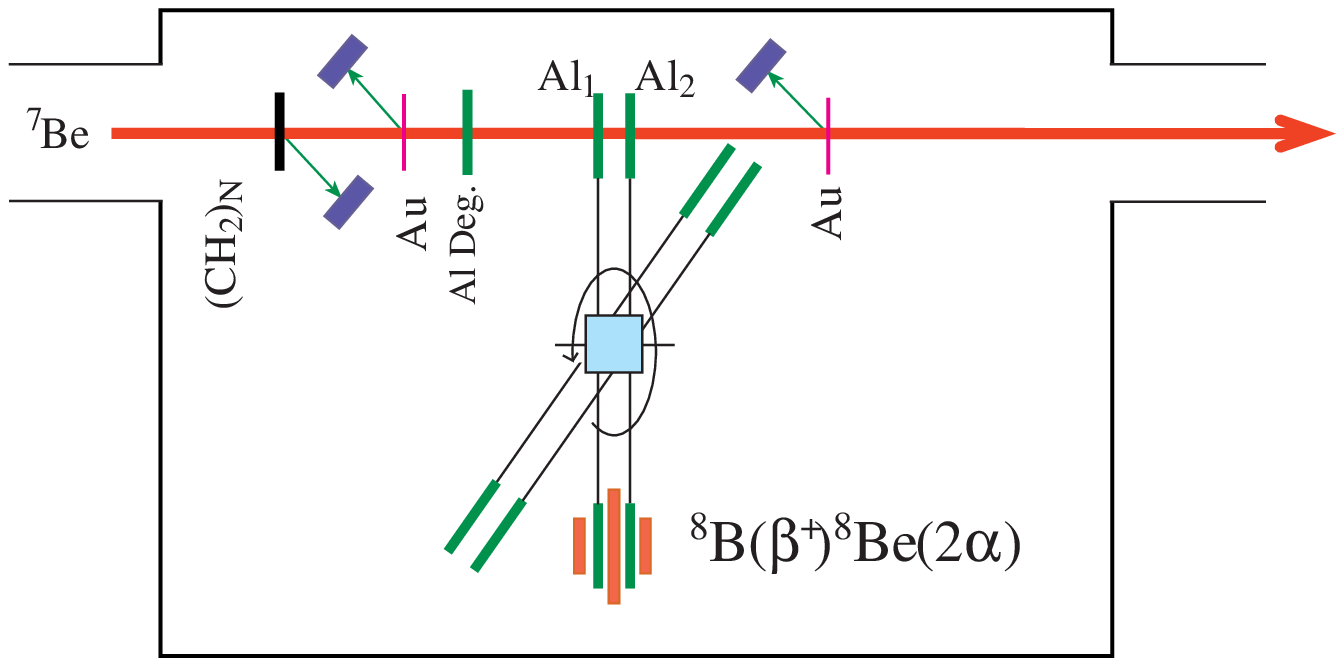,height=3in}}

\hspace{0.5in} \underline{Fig. 4:}  The Experimental setup of the UConn-LLN 
   Experiment (PH120) to measure the \be7pg reaction with \xbe7 beams 
   \cite{Gai98}.

The experimental detector setup for the LLN experiment is shown in 
Fig. 4. By measuring the energy
of the emerging \xbe7 beam, we ensure 
that the \b8 produced in the middle of the target is stopped at the end of
the first aluminum catcher foil. Due to the finite
target thickness, the recoil \b8 emerge with a (step) distribution of
energies with widths approximately 0.7 MeV,
and a stopping spread in aluminum of
approximately 0.5 $\mu m$. Thus the stopped \b8 are designed
to be equally spread over the two aluminum
catcher foils (0.5 $\mu m$ each).
The beta-delayed alpha-particle emission
of \b8 will be measured by measuring coincidence between
the two back to back equal energy alpha-particles detected in
a pair of detectors, see Fig. 4. The large diameter of 
the central detector (more than twice
that of front and back detectors) yield a large coincidence efficiency
(better than 98\%), as shown by our extensive Monte Carlo simulations.
By requiring that the count rates in the front and back
pair of detectors are similar we will ensure that the \b8 are spread
equally on both aluminum foils and thus no \b8 escapes the catcher foils
and none get stopped in the degrader, yielding nearly 100\%
collection efficiency.

In the target region,
two monitors measure beam intensity by measuring
the elastic scattering off a thin Au foil (evaporated onto
a very thin carbon backing) and the recoil
protons off the target. We plan to measure $\sigma / \sigma_{Rutherford}$
for p + \xbe7 with high precision. Thus the cross
section of the \be7pg reaction will
be measured relative to the elastic scattering, thereby removing several
systematic uncertainties related to beam-target composition.
The hydrogen component of the target will be
continuously monitored by measuring the recoil protons from the target.
These measurements will allow us to continuously measure and monitor
the beam-target luminosity.

Since two alpha-particles are associated with that decay we expect a very
large detection efficiency, approximately 50\% of $2\pi$.  Our
extensive Monte carlo simulations yield a large (98\%)
coincidence efficiency
and thus approximately 50\% total coincidence efficiency for two equal energy
correlated back to back
alpha-particles. For a \b8 transfer time of 0.07 sec, every 0.5 sec, we
obtain a total alpha-particle detection efficiency of approximately 25\%.
The closed detection geometry (50\% of 4$\pi$) with
a front and back detectors (a-la calorimetry
style) also ensures that the total alpha detection efficiency is nearly
independent of the exact location of the collection foils, as long as the
two foils remain parallel and at constant distance and the recoil \b8 nuclei
are spread equally on both catcher foils.

A beam intensity of $5 \times 10^8 \ /sec$ and a 250
$\mu g/cm^2\ CH_2$ target ($\Delta E_{cm}=100\ keV$)
containing $2 \times 10^{19}\ hydrogens/cm^2$
yield a luminosity of $10^{28}\ /sec/cm^2$. With expected cross sections
of $\sigma \ = \ 0.5,\ 0.4$ and $0.2 \ \mu b$, at
$E_{cm} \ = \ 1.0, \ 0.8$ and $0.5$ MeV, respectively,
and alpha-particle detection efficiency of 25\%, we obtain count rates
of approximately 5, 4, and 2 counts per hour. Thus experiments
lasting two to three days at
$E_{cm} \ = \ 1.0, \ 0.8$ and $0.5$ MeV, respectively, will yield a
total count of 240, 192 and 144 counts and statistical uncertainties
of 6.4\%, 7.2\% and 8.3\%, respectively. With approved 9 days of
experiment we plan to adjust the length of runs to achieve 5\% precision at
each data point.

\section{Acknowledgement} 

I would like to acknowledge the work of N. Iwasa, T. Kikuchi,
K. Suemmerer, F. Boue and P. Senger on the data analyses of the CD 
data and Ralph H. France III and James E. McDonald on the data 
analysis of the \be7pg reaction.  I also acknowledge
discussions and encouragements from
Professors J.N. Bahcall, C.A. Bertulani,
G. Baur, and Th. Delbar.


\begin{thebibliography}{9}

\bibitem{Book} John N. Bahcall, Neutrino Astrophysics,
Cambridge University Press, New York, 1989.

\bibitem{Phs96} J.N. Bahcall, F. Calaprice, A.B. McDonald,
       and Y. Totsuka; Physics Today{\bf 30,\#7}(1996)30 and references
therein.

\bibitem{Wo78} L. Wolfenstein, Phys. Rev. {\bf D17}(1978)2369,
          ibid {\bf D20}(1979)2634.

\bibitem{Mi85} S.P. Mikheyev, and A.Yu. Smirnov, Yad. Fiz.
       {\bf 44}(1985)847.

\bibitem{Ade98} E.G. Adelberger {\em et al.};
                 Rev. Mod. Phys. {\bf 70}(1998)1265.

\bibitem{Kavanagh60}
R.W.~Kavanagh, Nucl. Phys. {\bf 15}(1960)411.

\bibitem{Parker68}
P.D.~Parker {\it et al.}, Astrophys. J. {\bf 153}(1968)L85.

\bibitem{Kavanagh69}
R.W.~Kavanagh {\it et al.}, Bull. Am. Phys. Soc. {\bf 14}(1969)1209.

\bibitem{Vaughn70}
F.J.~Vaughn {\it et al.}, Phys. Rev. C {\bf 2}(1970)1657.

\bibitem{Wiezorek77}
C.~Wiezorek {\it et al.}, Z. Phys. A {\bf 282}(1977)121.

\bibitem{Filippone83}
B. W.~Filippone {\it et al.}, Phys. Rev. C {\bf 28}(1983)2222.

\bibitem{Hammache98}
F.~Hammache {\it et al.}, Phys. Rev. Lett. {\bf 80}(1998)928.

\bibitem{Ha99}	M. Hass et al., Phys. Lett. B462(1999)237.

\bibitem{Iw99} N. Iwasa {\em et al.}; Phys. Rev. Lett. {\bf 83}(1999)2910

\bibitem{Johnson92}
C.W.~Johnson {\it et al.}, Astrophys. J. {\bf 392}(1992)320.

\bibitem{Jennings98}
B.K.~Jennings, S.~Karataglidis, and T.D.~Shoppa;
   Phys. Rev. C {\bf 58} (1998) 3711

\bibitem{Bahcall92}
J.N.~Bahcall and M.H.~Pinsonneault, Rev. Mod. Phys. {\bf 64}(1992)885.

\bibitem{Ba87} C.A. Barnes {\em et al.}; Phys. Lett. {\bf 197}(1987)315.

\bibitem{Bau86} G. Baur, C.A. Bertulani, and H. Rebel;
 Nucl. Phys. {\bf A458}(1986)188.

\bibitem{Be88}  C.A. Bertulani and G. Baur; Phys. Rep. {\bf 163}(1988)299.

\bibitem{Pr51}  H. Primakoff; Phys. Rev. {\bf 81}(1951)899.

\bibitem{Ber94} C.A. Bertulani; Phys. Rev. {\bf C49}(1994)2688.

\bibitem{Typ94} S. Typel and G. Baur; Phys. Rev. {\bf C50}(1994)2104.

\bibitem{Ga95}  Moshe Gai, Nucl. Phys. {\bf B(Sup.)38}(1995)77.

\bibitem{Ki97} T. Kikuchi {\em et al.};
         Phys. Lett. {\bf B391}(1996)261.

\bibitem{Ki98} T. Kikuchi {\em et al.};
        Eur. Phys. J. {\bf A3}(1998)213.

\bibitem{Mo94}  T. Motobayashi {\em et al.};
        Phys. rev. Lett. {\bf 73}(1994)2680. and N. Iwasa {\em et al.};
        Jour. Phys. Soc. Jap. {\bf 65}(1996)1256.

\bibitem{Ga94} Moshe Gai, and Carlos A. Bertulani;
        Phys. Rev. {\bf C52}(1995)1706.

\bibitem{Davids} B.~Davids {\it et al.}, Phys. Rev. Lett. {\bf 81}(1998)2209.

\bibitem{Esben} H. Esbensen and G.F. Bertsch;
   Phys. Lett. {\bf B359}(1995)13 ibid
   531  Nucl. Phys. {\bf A600}(1996)37.

\bibitem{Ber98} Carlos A. Bertulani and Moshe Gai; Nucl. Phys.
  {\bf A636(1998)227}.

\bibitem{Senger93}
P. Senger {\it et al.}, Nucl. Instr. Meth. A {\bf 327}, 393 (1993).

\bibitem{Geissel92}
H. Geissel {\it et al.}, Nucl. Instr. Meth. B {\bf 70}, 286 (1992).

\bibitem{GEANT}
Detector description and simulation tool by CERN, Geneva, Switzerland.

\bibitem{Gai98} M. Gai, J.E. McDonald, R.H. France III, J.S. Schweitzer, 
 C. Angulo, Ch. Barue, S. Cherubini, M. Cogneau, Th. Delbar, M. Gaelens, 
 P. Leleux, M. Loiselet, A. Ninane, G. Ryckewaert, K.B. Swartz, D. Visser;
  	Bull. Amer. Phys. Soc. {\bf 44,II}(1999)1529.

\end{thebibliography}
\end{document}